\documentclass[aps,prl,superscriptaddress,groupedaddress,nofootinbib]{revtex4}  

\usepackage{graphicx}  
\usepackage{dcolumn}   
\usepackage{bm}        
\usepackage{amssymb}   
\usepackage{amsmath}

\def\be{\begin{equation}}
\def\ee{\end{equation}}
\def\bea{\begin{eqnarray}}
\def\eea{\end{eqnarray}}

\begin{document}

\title{Preheating with Fractional Powers}

\author{Hossein Bazrafshan Moghaddam, Robert Brandenberger\\
Department of Physics, McGill University, Montreal, QC H3A 2T8 Canada}

\begin{abstract}

We consider preheating in models in which the potential for the
inflaton is given by a fractional power, as is the case in
axion monodromy inflation. We  
assume a standard coupling between the inflaton field and a scalar matter
field. We find that in spite of the fact that the oscillation of the inflaton 
about the field value which minimizes the potential is anharmonic,
there is nevertheless a parametric resonance instability, and we
determine the Floquet exponent which describes this instability
as a function of the parameters of the inflaton potential.

\end{abstract}

\maketitle

\section{Introduction}

There has been a lot of recent interest in large field models
of inflation in which the potential is given by a fractional
power of the field. An example which is currently attracting
much attention is axion monodromy
inflation (see \cite{Eva} for the initial paper
and \cite{EvaReview} for a recent review).
Axion monodromy models are attractive since they may
provide a natural realization of large field inflation in
the context of superstring theory. Large field models
of inflation \cite{Starob, Linde} are advantageous since in such
a context the slow-roll trajectory is a local attractor in
initial condition space \cite{Kung}, even including metric
fluctuations \cite{Feldman}. In contrast, for small field
models the initial velocity of the inflaton field needs
to be fine tuned, thus creating a potential initial condition
problem \cite{Piran}. As is well known \cite{Lyth}, from 
the point of view of observations, large field models of
inflation are interesting since they may lead to a significant
tensor to scalar ratio. 

There are challenges to obtain large field inflation models.
For field values $|\phi| > m_{pl}$ corresponding to large
field inflation ($m_{pl}$ is the Planck mass), there is the
danger that gravitational corrections will lift the potential
and prevent slow rolling of the field, unless there is
a symmetry such as shift symmetry \cite{Masahide} which
protects the small mass required for large field inflation.
In string theory, there is a further challenge of obtaining
large field inflation: we expect the field range of the candidate
inflatons, e.g. the moduli fields or the fields associated
with brane separations, to be small, and hence incompatible
with large field inflation. Monodromy inflation \cite{Eva0}
provides a possible resolution of this problem, and axion
monodromy is currently regarded as the most promising
implementation of the idea of monodromy inflation in the
context of string theory \cite{EvaReview}. For this
reason, there has been a lot of recent activity on
this topic (see e.g. \cite{Dong-h,Conlon-b,Harigaya-i,Marchesano-s,Kappl-k,Arends-h-hk-l-m-s-w,
McAllister-w-w,Franco-g-r-u,Ben-Dayan-p-w,Blumenhagen-h-p,Higaki-t,Bachlechner-l} 
for a selection
of recent papers).

To our knowledge there has been very little work on
reheating in models in which the inflaton potential
is given by a fractional power of the field $\Phi$ (but see
\cite{Knauf} for a study of reheating 
in the context of the initial brane monodromy
inflation model of \cite{Eva0}). 
The goal of this paper is to consider preheating
in models in which the inflaton potential is given by a 
fractional power of the field, and in which
the inflaton field is coupled to a second scalar matter
field $\chi$ via a usual interaction term $\Phi^2 \chi^2$ \footnote{For a perturbative study in the context of axion
monodromy inflation see \cite{Blumenhagen}.}.Our goal here is to study the
changes in preheating which arise due to the fact that the oscillations of the 
inflaton about the field value which minimizes the potential are not
harmonic. 

Reheating is an integral part of any inflationary model.
Without reheating the universe after inflation would be
empty and cold. Reheating was initially discussed using
first order perturbation theory \cite{original,Dolgov-l,Albrecht-s-t-w}. In
\cite{TB} and \cite{DK} it was realized that an oscillating
inflaton field will induce a parametric resonance instability
in fields which couple to the inflaton (and also in the
equations for the inflaton fluctuations themselves). This
instability was then worked out in detail in \cite{KLS1, STB, KLS2}
and given the name ``preheating''\footnote{For recent
reviews of reheating the reader is referred to \cite{ABCM}
and \cite{Johanna}.}. 

Reheating has been analyzed mostly for inflaton potentials
of the form $V(\phi) = \frac{1}{2} m^2 \phi^2$ and 
$V(\phi) = \lambda \phi^4$, and for a matter field $\chi$
(treated for simplicity as a scalar field) coupled to $\phi$
via an interaction Lagrangian ${\cal{L}}_{int} \sim \phi^2 \chi^2$.
For both types of inflaton potential, the
linearized equation of motion for fluctuations of
$\chi$ leads (neglecting the expansion of the universe) to
a harmonic oscillator equation with a periodically varying
mass. In the first case of a massive inflaton field, the
equation for Fourier modes of $\chi$ is
an equation of Mathieu type \cite{Mathieu}, in the 
second case, the case of a massless inflaton field with
quartic potential, to a generalized equation. In both
cases, the equation of motion for the Fourier modes
$\chi_k$ is of Floquet type, and the general theory
\cite{Floquet} tells us that the modes will evolve as
\be
\chi_k \, \sim \, P_{1,k}(t) e^{\mu_k t} + P_{2, k}(t) e^{- \mu_k t}\, ,
\ee
where $P_{i,k}(t)$ $(i = 1, 2)$ are periodic functions, 
and $\mu_k$ is a non-negative
exponent called the ``Floquet exponent''.
Values of $k$ for which $\mu_k > 0$ form resonance
bands.

Past work on reheating has shown that the efficiency
of the parametric resonance instability depends
on which class of model one is considering. We are interested in
inflaton potentials with fractional power of the field. As in the
canonical inflation models, the inflaton will be oscillating
about its ground state. However, the oscillation is not
sinusoidal. Given the recent
interest in axion monodromy inflation it is hence of great
interest to study the efficiency of reheating. This is the
problem we address in this paper. We will study the
structure of the resonance bands and estimate the
value of the Floquet exponent within the resonance bands.
Our results show that the parametric instability is of
broad resonance type and hence efficient. It will lead
to the transfer of the inflaton energy density from the
homogeneous inflaton background to matter fluctuations
within less than a Hubble expansion time. Our result
will allow a more precise determination of the parameters
of models such as axion monodromy inflation from
cosmological observations. As has been emphasized
recently (see e.g. \cite{why,Munoz-k}), for a given action the
values of the cosmological parameters such as the
scalar spectral index $n_s$ and the tensor to scalar
ratio $r$ depend on the reheating history. One of the
reasons for this is that changing the duration of reheating
will change the current physical wavelength which
exits the Hubble radius a fixed number of e-foldings
before the end of inflation. Our result also has implications
for baryogenesis. Since the energy transfer to regular
matter after inflation is rapid, the ``reheating temperature"
is high and high-scale baryogenesis mechanisms become
feasible.

In the next section we give a brief review of axion monodromy
inflation as an example of the type of models we have in
mind. Section 3 is the main one in which we study
the equation of motion for the fluctuations of a matter
field coupled to the inflaton and establish that 
preheating is efficient. We work in the approximation
that the expansion of space is negligible. Since the
instability in our model is of broad resonance type and
not of narrow resonance, including effects of the
expanding space will not change the qualitative
features of our results. I
The most important change is that instead of exponential growth
of the fluctuations we will have exponential growth
damped by a power of the cosmological scale factor
In Section 4 we show that this approximation is self-consistent and
we analyze various back-reaction effects. 

We use standard notation in which $a(t)$ is the
cosmological scale factor, $t$ is physical time,
and $H(t)$ is the Hubble expansion rate.

\section{Axion Monodromy Inflation}

The low energy limit of string theory provides many
scalar fields which can be candidates for an inflaton
field (see e.g. \cite{StringInflation} for a recent
comprehensive review of inflation in the context of
string theory). Examples are radii of extra dimensions
or distances between brane-antibrane pairs. Typically,
however, the range of field values is small and does
not admit super-Planck values required for large-field
inflation. 

Monodromy \cite{Eva0} is a simple solution to this
problem. A string connecting brane-antibrane pairs
can be wrapped many times about an internal cycle,
thus yielding large field values even if the basic
field range is small. As was shown in \cite{Eva},
suitably wrapped branes induce trans-Planckian field
ranges for the associated axion fields. In string
compactifications, axions arise from integrating
gauge potentials over cycles in the internal manifold.
The potential for the axions induced by the presence
of the branes of the axion field is not periodic but 
grows without bound, however with a power which is 
typically less than $2$ \footnote{Very close to the
origin the inflaton potential can be quadratic \cite{Eva0},
but that field range is negligible in terms of setting
up the overall dynamics of $\phi$. As we will see,
what matters for the efficiency of preheating is the
amplitude of ${\dot{\phi}}$ near the minimum
of the potential, and this is set by the asymptotics
of the potential for field values where inflation
ends. We thank E. McDonough for alerting us to
this issue.} . One example studied in \cite{Eva} has a
linear potential, but potentials with fractional powers
$p$, e.g. $p = 2/3$, also arise (see e.g. \cite{Eva0}). The potentials
have oscillatory terms (caused by 
instanton effects) and thus can give interesting observational
signals in the spectrum of cosmic microwave background
anisotropies (see e.g. \cite{Elisa,p-e-f,e-f,k-s-y,h-k-s-y,f-c-s-w,cai-f-q,amin11,zhou1}). However, for 
powers $p < 2$ which we consider here the oscillatory 
terms are negligible for field values close to
the minimum of the potential (which we take to
be at $\phi = 0$). It is the dynamics near 
$\phi = 0$ which determines the efficiency of
reheating, and hence the oscillating terms in
the potential will have a negligible effects
on the reheating efficiency. We will hence here consider
simple power law potentials.

\section{Evolution of Fluctuations after Inflation}

As discussed in previous section, we will consider the effective
potential for inflaton field to be 
\be \label{Vpot}
V_{eff}(\phi) \, = \, m^{4-p}\left|\phi\right|^{p} \, ,
\ee
where $m$ is a parameter with units of mass. 
We will also assume a simple coupling of the form 
\be
{\cal{L}}_{int} \, = \, \frac{1}{2}g^{2}\phi^{2}\chi^{2}
\ee
between the inflaton $\phi$ and a 
matter field $\chi$. Such a coupling arises at tree level
(but at second order in the axion coupling constant) between
the axion and gauge fields (which we can then represent
in a simplifying way by a scalar matter field $\chi$).
Therefore in the rest of the paper we will work with 
the potential
\begin{equation}
V(\phi,\chi) \, = \, m^{4-p}\left|\phi\right|^{p}
+ \frac{1}{2}g^{2}\phi^{2}\chi^{2} \, .
\end{equation}

If we take the matter field $\chi$ to be massless on
the scale of inflation, and if we neglect self-interactions
of $\chi$, then each Fourier mode $\chi_k$ of $\chi$
will evolve independently with an equation of motion
\begin{equation} \label{modeeq1}
\ddot{\chi}_{k} + 3H\dot{\chi}_{k} +
(\frac{k^{2}}{a(t)^{2}} + g^{2}\phi(t)^{2})\chi_{k} \, = \, 0 \, .
\end{equation}
This differential equation has a time-dependent frequency
and a gravitational damping term. The time dependence is 
given by evolution of both the scale factor $a(t)$ and of
the background inflaton field $\phi(t)$. 

Following \cite{TB}, we will neglect the expansion of the 
universe for the moment, and we will later show that
this is a self-consistent approximation during the early 
stages of preheating) and thus take $a(t)=1$ \footnote{The
expansion of space can be taken into account using the
methods of \cite{STB, KLS2}.}. Then we can rewrite 
Eq. (\ref{modeeq1}) as
\begin{equation} \label{modeeq2}
\ddot{\chi}_{k} + f(t)^{2} \chi_{k} \, = \, 0 \, ,
\end{equation}
where
\begin{equation} \label{fformula}
f(t)^{2} \, \equiv \, k^{2} + g^{2}\phi(t)^{2} \, .
\end{equation}

If $f^{2}(t)$ is a periodic function of time, then the equation
is of Floquet type \cite{Mathieu, Floquet} and we hence there
is the possibility of a parametric resonance instability for the 
$\chi$ fluctuations. The strength of the resonance depends
sensitively on the dynamics of $\phi$ near the minimum of the
potential. We will consider the case in which the homogeneous 
value of the $\chi$ field is zero. Since the
effective potential for the inflaton in this case is given by (\ref{Vpot}), 
the equation of motion for the background inflaton field is
\begin{equation} \label{inflatoneq}
\ddot{\phi} + 3H\dot{\phi} + pm^{4-p}\left|\phi\right|^{p-2}\phi \, = \, 0 \, .
\end{equation}
As mentioned above, we will neglect the expansion of space
and later check for the self consistency of this approximation. In
this case, the equation becomes
\begin{equation}
\ddot{\phi} + pm^{4-p}\left|\phi\right|^{p-2}\phi \, = \, 0 \, .
\end{equation}
We consider values $0 < p \leqslant 2$ for the exponent in
the potential. In spite of the fact that this equation is nonlinear
and does not describe harmonic oscillation of inflaton field,
the evolution of the inflaton field is periodic (periodic motion
about the minimum of the potential). 

As we know from previous studies of preheating \cite{KLS2},
the efficiency of preheating is determined by the dynamics
of $\phi$ near the minimum of its potential, i.e. by the
value of ${\dot \phi}$ at the time $t_c$ when $\phi = 0$.
This value can be determined by energy conservation. The
energy density at the end of inflation $(t = t_{end})$ is given by
\begin{equation}
\rho_{end} \, = \, \frac{1}{2}\dot{\phi}^{2}(t_{end}) +
m^{4-p}\left|\phi\right|^{p}(t_{end})
\end{equation}
from which we can find
\begin{equation}
\dot{\phi}(t_c) \, = \, \sqrt{2\rho_{end}} \, .
\end{equation}

Inflation ends when the slow-roll conditions break down. These are
\bea
\frac{1}{2} {\dot{\phi}}^2 \, &<& \, V(\phi) \,\,\, {\rm{and}} \\
3 H {\dot{\phi}} \, &<& \, V^{\prime} \, ,
\eea
where the prime denotes the derivative with respect to $\phi$. The
first condition breaks down once
\be \label{phien}
|\phi| \, = \, \frac{p}{\sqrt{48 \pi}} m_{pl} \, .
\ee
For $1 \leq p \leq 2$ the second condition breaks down for a smaller
value of $|\phi|$. For $0 < p < 1$ the number $p$ in the numerator
of (\ref{phien}) is replaced by $\sqrt{p(2 - p)}$. We will denote
the end point of inflation by $\phi_e$ and will write
\be \label{phiend}
\phi_e \, \equiv \, \alpha m_{pl} \, 
\ee
where the value of the constant $\alpha$ can be read off from 
(\ref{phien}), and we have used $G \equiv m_{pl}^{-2}$. It is
convenient to parametrize the end point of inflation in terms
of a dimensionless number $\sigma$ given by
\be \label{sigmadef}
\sigma m \, \equiv \, \alpha m_{pl} \, .
\ee
This will simplify some of the later equations.

Now we turn to the study of $\chi$ particle production, i.e. to the study of 
the structure of the resonance in the Eq.  (\ref{modeeq2}). While the
the adiabaticity condition
\be \label{adiab}
\frac{df(t)}{dt} \, < \, f(t)^{2} 
\ee
is satisfied then using the WKB approximation we conclude that the state tracks
the instantaneous vacuum and we find the solution of the from
\begin{equation} \label{adsol}
\chi_{k}=\frac{\alpha_{k}}{\sqrt{2f(t)}}\exp(-i\int f(t)dt)+\frac{\beta_{k}}{\sqrt{2f(t)}}\exp(i\int f(t)dt)
\, ,
\end{equation}
where $\alpha_k$ and $\beta_k$ are coefficients determined by the initial
conditions.

For small field values, the adiabaticity condition (\ref{adiab}) is
violated. To estimate the range of field values for which this
is the case, we can replace ${\dot{\phi}}(t)$ by ${\dot{\phi}}(t_c)$,
namely
\begin{equation} \label{vel}
\dot{\phi} \, \simeq \, 2 \sqrt{m^{4-p}(\alpha M_{pl})^{p}} \, = \, 2 \sigma^{p/2} m^2 \, ,
\end{equation}
since for small field values ${\dot{\phi}}$ is constant to a good
approximation. Making use of the formula (\ref{fformula}) for $f(t)$ 
we see that adiabaticity violation arises for field values
obeying
\be
2 g^{2} \sigma^{\frac{p}{2}} m^{2} \phi \, 
\geq \, \bigl( k^2 + g^{2}\phi^{2} \bigr)^{3/2} \, .
\ee
Considering modes in the far infrared (i.e. setting $k = 0$) we
find 
\begin{equation}
\left|\phi\right| \, \leq \, \frac{\sqrt{2} \sigma^{\frac{p}{4}}m}{\sqrt{g}} \, .
\end{equation}
We denote the resulting value of $|\phi|$ by $\phi_{max}$. 
Thus we conclude that for small values of $k$ the condition for violation
of adiabaticity is
\begin{equation} \label{range}
-\frac{\sqrt{2} \sigma^{\frac{p}{4}}m}{\sqrt{g}} \, \leq \, \phi \, \leq \,
\frac{\sqrt{2}\sigma^{\frac{p}{4}}m}{\sqrt{g}} \, .
\end{equation}

In order to estimate the energy transfer via this parametric resonance
instability it is important to estimate the width of the instability band. To
a first order, this is given by the range of $k$ values for which
the approximation made above of neglecting $k$ in the adiabaticity
condition violation calculation is self-consistent, namely value of $k$
for which $k < g \phi_{max}$. This gives
\be \label{kmax}
k_{max} \, = \, \sqrt{2g} \sigma^{p/4} m \, .
\ee
To put this value in context, note that for $p < 4$ this value is parametrically
larger than $H$, which means that the modes which dominate
the phase space of modes which undergo parametric resonance have 
wavelength smaller than the Hubble radius $H^{-1}$ at the end of inflation,
and that hence neglecting metric fluctuations is well justified (see
\cite{RHBfluctreview} for a discussion showing that metric fluctuations
have a negligible effect if the wavelength is smaller than the Hubble
radius). Note that even modes with larger values of $k$ feel
the resonance, but for a shorter time interval, namely a time interval
corresponding to
\be
|\phi| \, < \, g^{-1} k \, .
\ee

Now we return to Eq. (\ref{modeeq2}) and want to find amplification of $\chi_{k}$
modes after passing once through the range of $\phi$ which allows resonance.
Using Eq. (\ref{vel}) for the range (\ref{range}) of field values where resonance 
can take place we can write (by rescaling time)
\begin{equation}
\phi \, \simeq \, 2 \sigma^{\frac{p}{2}}m^{2}t \, .
\end{equation}
Rewriting Eq. (\ref{modeeq2}) in terms of new time variable
\be
\tau \, \equiv  \, (2 g)^{\frac{1}{2}}\sigma{}^{\frac{p}{4}}mt
\ee
we get
\begin{equation} \label{modeeq3}
\chi_{k}'' + (q^{2} + \tau^{2})\chi_{k} \, = \, 0 \, ,
\end{equation}
where the rescaled wavenumber $q$ is
\be
q \, \equiv \, \frac{k}{(2g)^{\frac{1}{2}}\sigma{}^{\frac{p}{4}} m} \, .
\ee

Equation (\ref{modeeq3}) describes scattering from a 
negative parabolic potential in quantum mechanics. According to
the usual analogy where the classical variable $\chi$
becomes the quantum mechanical wave function, and the
time variable in classical dynamics becomes the spatial variable
in the quantum problem, the equation (\ref{modeeq3}) 
corresponds to the time-independent  Schr\"odinger equation 
\be
\nabla^{2}\psi + (E-V)\psi \, = \, 0 \, ,
\ee
for a wave function $\psi$ in a parabolic potential $V = \tau^2$ and with energy
$E$. This analogy has been used in the context of reheating
in \cite{KLS2}, and in the case of reheating in the presence of noise
in \cite{Craig,Z-m-z-b}. The time range in the reheating problem when the
field is in the range (\ref{range}) corresponds to values of $x$ where
quantum tunneling occurs. We will use the standard methods
from quantum mechanics to solve the problem.
 
Well before $\phi$ enters the range (\ref{range}) the solution is of the form 
(\ref{adsol}),namely  
\begin{equation} \label{adsol1}
\chi_{k}^{0} \, = \, \frac{\alpha_{k}^{0}}{\sqrt{2f(\tau)}}\exp(-i\int f(\tau)d\tau)+\frac{\beta_{k}^{0}}{\sqrt{2f(\tau)}}\exp(i\int f(\tau)d\tau) \, .
\end{equation}
Long after it passes through that region, the solution again 
is of the same form, namely 
\begin{equation} \label{adsol2}
\chi_{k}^{1} \, = \, \frac{\alpha_{k}^{1}}{\sqrt{2f(\tau)}}\exp(-i\int f(\tau)d\tau)+\frac{\beta_{k}^{1}}{\sqrt{2f(\tau)}}\exp(i\int f(\tau)d\tau) \, .
\end{equation}
The coefficients in Eq. (\ref{adsol1}) come from initial conditions.
We can find the coefficients in Eq. (\ref{adsol2}) using amplitudes of reflection
$R_{k}$ and transmission $D_{k}$ for the scattering. These coefficients
are given by \cite{KLS2}
\begin{equation}
R_{k} \, = \, -\frac{ie^{i\varphi_{k}}}{\sqrt{1+e^{\pi q^{2}}}}
\end{equation}
and
\begin{equation}
D_{k} \, = \, \frac{e^{-i\varphi_{k}}}{\sqrt{1+e^{-\pi q^{2}}}} \, ,
\end{equation}
where 
\be
\left|R_{k}\right|^{2} + \left|D_{k}\right|^{2} \, = \, 1
\ee
and $\varphi_{k}$ is given by
\begin{equation}
\varphi_{k} \, = \, \arg\Gamma(\frac{1+iq^{2}}{2})+\frac{q^{2}}{2}(1+\ln\frac{2}{q^{2}}) \, .
\end{equation}
We also note that the dependence on the phase of incoming wave during resonance
the amplitude can increase or decrease. Making use of these results, and 
including the phase $\theta_{k}^{0}$ of the incoming wave in the 
calculation, we obtain the following matrix relation between the coefficients of
the positive and negative frequency modes before and after the scattering:
\begin{equation}
\begin{pmatrix}\alpha_{k}^{1}\\
\beta_{k}^{1}
\end{pmatrix}=\begin{pmatrix}\sqrt{1+e^{-\pi q^{2}}}e^{i\varphi_{k}} & ie^{-(\frac{\pi}{2})q^{2}+2i\theta_{k}^{0}}\\
-ie^{-(\frac{\pi}{2})q^{2}-2i\theta_{k}^{0}} & \sqrt{1+e^{-\pi q^{2}}}e^{-i\varphi_{k}}
\end{pmatrix}\begin{pmatrix}\alpha_{k}^{0}\\
\beta_{k}^{0} 
\end{pmatrix}
\end{equation}

Returning to our reheating problem, we see that $\left|\beta_{k}^{1}\right|^{2}$ 
gives the occupation number of particles with momentum
$k$ after one passage of $\phi$ through the minimum of its potential, given
that the initial occupation number is $\left|\beta_{k}^{0}\right|^{2}$.
Considering $n_{k}^{1}=\left|\beta_{k}^{1}\right|^{2}$ and the same
relation for incoming particles we get
\begin{equation} \label{result1}
n_{k}^{1} \, = \, \exp(-\pi q^{2})+(1+2\exp(-\pi q^{2}))n_{k}^{0}-2\exp(-\frac{\pi}{2}q^{2})\sqrt{1+\exp(-\pi q^{2})}\sqrt{n_{k}^{0}(1+n_{k}^{0})}\sin(\theta_{tot}^{0})
\end{equation}
where 
\be
\theta_{tot}^{0}=2\theta_{k}^{0}-\varphi_{k}+\arg\beta_{k}^{0}-\arg\alpha_{k}^{0} \, .
\ee
The above relation between initial number density and final number density
can be applied in sequence to all time intervals during which the adiabaticity
condition is violated. Therefore, for the j'th scattering event we obtain 
(in the case of large occupation number)
\begin{equation}
n_{k}^{j+1} \, \sim \, 
(1+2\exp(-\pi q^{2})-2\sin(\theta_{tot}^{j})\exp(-\frac{\pi}{2}q^{2})\sqrt{1+\exp(-\pi q^{2})})n_{k}^{j} \, .
\end{equation}
Defining the ``Floquet exponent'' $\mu_{k}^{j}$ for the j'th scattering through
\be
n_{k}^{j+1} \, = \, \exp(2\pi\mu_{k}^{j})n_{k}^{j}
\ee
we obtain
\begin{equation} \label{FE}
\mu_{k}^{j} \, = \, \frac{1}{2\pi}\ln(1+2\exp(-\pi q^{2})-2\sin(\theta_{tot}^{\text{j}})\exp(-\frac{\pi}{2}q^{2})\sqrt{1+\exp(-\pi q^{2})}) \, .
\end{equation}
In Figure 1 we sketch the value of $\mu_{max}$ as a function of $k$.

\begin{figure}
  \includegraphics[width=0.6\linewidth]{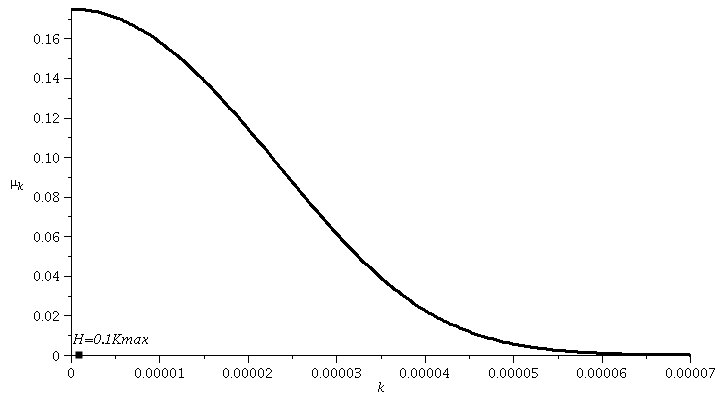}
  \caption{Sketch of the Floquet exponent from Eq. (41) (vertical axis)
as a function of the value of $k$ (horizontal axis) for typical value of $\sin{\theta}=0$, $p=1$, $\alpha=0.1$, $m=10^{-4}m_p$ and $g=10^{-3}$. The values
on the horizontal axis are in Planck units. We also mark the
value of $H$ on this axis.}
 
\end{figure}

Summing over all scattering events,
considering the initial conditions
$n_{k}^{0}=0$, $\alpha_{k}^{0}=1$,$\beta_{k}^{0}=0$, 
and taking the phases $\theta_{k}^{j}$ to be randomly 
distributed we obtain
\begin{equation}
n_{k}(t) \, = \, \frac{1}{2}\exp(2\pi\sum_{j}\mu_{k}^{j})
\end{equation}
We can rewrite the exponent as 
\begin{equation}
\sum_{j}\mu_{k}^{j} \, \equiv \, \mu_{eff}N
\end{equation}
where $N$ is the number of oscillations of inflaton field. 
At a time $t$ after the end of inflation, the number $N$
is given by
\begin{equation}
N \, = \, \frac{t}{T/2}
\end{equation}
where $T$ is the period of infalton field oscillations is $T$ and 
we devided by two since in each period of oscillation, the
inflaton field passes through the non-adiabatic region two times. 

The period $T$ can be estimated taking $\dot{\phi}$ to be the
constant value given in (\ref{vel}) and solving
\be
T {\dot{\phi}} \, = \, 4 \sigma m \, ,
\ee
since one quarter of the integrated field range over one
period is $\sigma m$. Hence
\begin{equation}
T \, \sim \, \frac{2 \sigma^{1-\frac{p}{2}}}{m} \, .
\end{equation}
Therefore 
\begin{equation}
\sum_{j}\mu_{k}^{j} \, \sim \, \mu_{eff}\frac{mt}{\sigma^{1-\frac{p}{2}}} \, .
\end{equation}

The next step is to find the value for $\mu_{eff}$ . Considering
small values for $q$ and averaging over $\sin\theta_{tot}$, we
find from (\ref{FE})
\be
\mu_{eff} \, \simeq \, \frac{1}{2\pi}\ln(3) \, .
\ee

To find the energy density $\rho_{\chi}$
of the particles produced during preheating
we need to integrate $n_{k} k$ over all momenta which experience
parametric resonance, i.e. with $|k| < k_{max}$ 
\bea
\rho_{\chi}(t) \, &=& \, \int_{0}^{k_{max}} d^{3}k' k' \frac{1}{2} 
\exp(2\pi\mu_{eff}\frac{mt}{\sigma^{1-\frac{p}{2}}}) \\
&\simeq& \, \frac{4\pi}{9} k_{max}^4\exp(\ln(3)\frac{mt}{\sigma^{1-\frac{p}{2}}})
\eea
where $k_{max}$ is given by (\ref{kmax}). Inserting the value for $k_{max}$ 
we find
\begin{equation} \label{dens1}
\rho_{\chi}(t) \, \simeq \, \frac{16\pi}{9} g^2 m^{4} \sigma^p exp(\ln(3)\frac{2mt}{\sigma^{1-\frac{p}{2}}}) \, ,
\end{equation}
which is to be compared with the inflaton density at the end of inflation which
is
\be \label{dens2}
\rho(t_e) \, \simeq 2 m^4 \sigma^p \, .
\ee
We conclude that provided that the resonance lasts so long that the exponential
growth factor can overcome the factor $\frac{16\pi}{9} g^2$ by which (\ref{dens1})
is suppressed compared to (\ref{dens2}), then the parametric resonance
instability will be sufficiently strong to drain all of the energy stored in the
inflaton field at the end of inflation. Whether this is the case or not depends
on how soon back-reaction effects take over and how soon the approximations
we have made cease to be self-consistent.

\section{Back-Reaction Effects}

In the analysis of the previous section we have neglected the expansion
of space. In toy models such as those studied in \cite{STB, KLS2} this
is a self-consistent approximation. We must check whether 
neglecting the expansion of space is justified in the present model.
First of all, note that  our
analysis shows that the non-adiabatic evolution of the solution
of the equation of motion for $\chi$ takes place in narrow time
intervals $\Delta t$ when $\phi$ is close to $\phi = 0$. Combining (\ref{vel}),
(\ref{range}) and the value for $H$ given by the potential energy
density at the end of inflation, i.e. at $\phi = \phi_e$ we
obtain
\be
\frac{H^{-1}}{\Delta t} \, \sim \, \frac{m_{pl}}{m} \, ,
\ee
and hence we conclude that the cosmological expansion does
not affect the study of $\chi$ particle production during the
resonant time interval about any one $\phi = 0$ crossing point.
Since the period $T$ of oscillation obeys
\be
T H^{-1}(t_e) \, \ll \, 1 \, ,
\ee
neglecting the expansion of space is also justified for any one
given oscillation period. Finally, taking as an example $p = 1$
and $m = 10^{-4} m_{pl}$, which yields a Hubble expansion
at the end of inflation close to the upper bound from gravitational
radiation constraints, we find
\be
\frac{m H^{-1}(t_e)}{\sigma^{1 - p/2}} \, \sim \alpha^{-1} \, \sim \, 10 \, .
\ee
Thus, over one Hubble expansion time the exponential increase in
$\rho_{\chi}$ from (\ref{dens1}) is sufficient to overcome the small
factor of $g^2$ in (\ref{dens1}), taking $g^4 \sim 10^{-12}$ such that
the corrections to the potential for $\phi$ due to $\chi$ loops does
not dominate over the original $V(\phi)$ at the field value $\phi \sim m_{pl}$.
Thus, we conclude that it is a reasonable approximation to neglect
the expansion of space.

We will now consider some back-reaction effects which could
terminate or even prevent preheating. First of all,
the fluctuations of $\chi$ which are generated in
the preheating instability will lead to correction
terms both in the equation of motion for $\phi$ and
in that of $\chi_k$. We need to determine the length
of time these effects can be neglected. Secondly, fluctuations
in the inflaton field themselves could be amplified and then
back-react both in the equation of motion for $\phi$, shutting
off the oscillations which drive preheating, and also in
the equation of motion for $\chi$, providing corrections to
the mass term which will prevent the violation of adiabaticity
which is required to obtain the resonance. 

We first discuss the possible amplification of fluctuations
of $\phi$. The equation of motion for the 
$\delta \phi_k$ mode is
\begin{equation}
 \ddot{\delta\phi_k} + 3H\dot{\delta\phi_k} +
[\frac{k^{2}}{a^{2}}+p(p-1)m^{4-p}\phi^{p-2}]\delta\phi_k \, = \, 0 \, .
\end{equation}
To solve this equation and study possible resonance effects, 
we can for the moment neglect the expansion of the universe.  
Since we are interested in large scale modes we can also drop 
the $k^{2}$ term.  Rewriting the equation and multiplying both sides 
by $\phi^{2-p}$ gives
\begin{equation}
 \phi^{2-p}\ddot{\delta\phi}  + p(p-1)m^{4-p}\delta\phi \, = \, 0 \, ,
\end{equation}
which (using (\ref{phiend}) and (\ref{sigmadef})) can be re-written as
\begin{equation} \label{phiflucteq}
 A_{1} t^{2-p}\ddot{\delta\phi} + A_2\delta\phi \, = \, 0 \, ,
\end{equation}
where 
\bea 
A_1 \, & \equiv & \, [ 2(\alpha m_{pl})^{\frac{p}{2}}m^{\frac{4-p}{2}}]^{2-p} \nonumber \\ 
A_2 \, & \equiv & \, p(p-1)m^{4-p} \, . 
\eea
The solutions of (\ref{phiflucteq}) are Bessel functions of the first 
and second kind and take the following form
\begin{equation}
 \delta\phi \,
  = \, c_1 \sqrt{t} J(\frac{1}{p}, \frac{2}{p} \sqrt{\frac{A_2}{A_1}} t^{p/2}) \,
 \,+ c_2 \sqrt{t} Y(\frac{1}{p}, \frac{2}{p} \sqrt{\frac{A_2}{A_1}} t^{p/2}) \,.
\end{equation}
The solutions for $\delta\phi$ have a mild time dependence (much weaker than
exponential). Hence, there is no parametric resonance of $\phi$ fluctuations,
and thus no important back-reaction effects of these modes.

Thus, we now turn to the discussion of the back-reaction effects of the
$\chi$ fluctuations. These fluctuations will lead to an induced potential
for the inflaton in Eq. (\ref{inflatoneq}), the magnitude of which is
given by $g^{2}\chi^{2}\phi^{2}$. This term will have a major effect on
inflaton dynamics when the interaction term become as important as the 
bare potential term $m^{4-p}\phi^p$. Therefore the criterion for this 
back-reaction effect to be negligible is
\begin{equation} \label{crit1}
\chi^{2}_{eff} \, \ll \, \frac{m^{4-p}}{g^2}\phi^{p-2} \, ,
\end{equation}
where the value of the inflaton field to be inserted is the one at the 
end of inflation.

Due to nonlinearities in the $\chi$ sector (which will inevitably be
present due to renormalization considerations) there is a term in the potential 
of the form $\frac{1}{4}\lambda\chi^4$ which will add a term to the mode
equation (\ref{modeeq1}). In the Hartree approximation, this term will
be an extra mass of magnitude $\lambda <\chi^2>$. Computing this effective 
mass by averaging over the modes which undergo parametric amplification, and
comparing with the amplitude of the mass term which is driving the
parametric resonance, we find that the induced mass term is negligible if
\begin{equation} \label{crit2}
\chi_{eff}^2 \, \ll \, \frac{g^{2}}{\lambda}\phi^{2} \, ,
\end{equation}
where once again the value of $\phi$ at the end of inflation is to
be used.

The observed amplitude of density fluctuations constrains the value of
the coupling constant $g^2$: at one loop level the interaction between
$\chi$ and $\phi$ induces a quartic self coupling term of $\phi$. The
coefficient of this term is proportional to $g^4$ but it is constrained to
be smaller than $10^{-12}$. Hence the constraint on $g^2$ is
$g^2 < 10^{-6}$. In turn, the same loop effects induce the self
coupling of $\chi$, and we expect $\lambda \sim g^4$. Since
the value of $m$ is also constrained to be small from the 
observed amplitude of the power spectrum of density fluctuations,
we expect the first criterion (\ref{crit1}) to be more stringent than
the second one (\ref{crit2}).

To see what the criterion (\ref{crit1}) implies, note that when
$\chi_{eff}^2$ reaches the limit given by (\ref{crit1}), then the
energy density in the $g^2 \phi^2 \chi^2$ term in the Lagrangian
is of comparable magnitude to the initial energy density in the
inflaton field, and (for values of $p < 2$) the energy density
from the nonlinear terms in the potential for $\chi$ is larger.
Thus, we conclude that the back-reaction effects studied here
do not terminate the preheating instability before a fraction
of order $1$ of the initial inflaton energy density has been
drained from the coherent oscillation of $\phi$. Hence, we
conclude that the parametric resonance instability is very
effective.

\section{Conclusions}

In this paper we have studied the effects on preheating which
result when considering the ``non-standard'' inflaton potentials
in which the potential scales as a fractional power of the
field, as is the case in
axion monodromy inflation. We have shown that the
parametric resonance instability is expected to be very
effective \footnote{While our work was being written up,
a paper \cite{Dolgov} appeared which also studied the
dependence of the efficiency of preheating on the shape
of the inflaton potential and concluded that changes from 
the canonical shape do not decrease the efficiency of
the loss of energy from the inflaton condensate.}. This
result will allow a more precise determination of cosmological
parameters in a given inflationary model of the type we
consider, and it also implies that high scale baryogenesis
scenarios are possible.

In this work we have used a standard $\phi^2 \chi^2$
coupling between the inflaton and the preheat fields. Whereas
such a coupling will arise at higher orders in the coupling
constant in actual axion monodromy inflation models, in such
models there are other couplings which are more important.
In work in progress we are studying preheating for these more
realistic couplings. In these more realistic models we can
then also study the possible preheating of entropy modes of
the metric fluctuations. Such metric instabilities could
lead to important constraints on axion monodromy inflation
models \cite{entropy,bbv,ffrhb,hbmrhbye}.

Our work has been based on analytical approximations
and we have not made use of the various numerical
codes which have been developed \cite{codes,fng,fva,sj,rfr,zh,sj2}. Since
we are interested in the onset and overall efficiency
of the parametric resonance instability, an analytical
approach is not only sufficient but maybe superior
since it provides analytical insight into why we obtain
the dynamics which we see.

\section{Acknowledgements}

We thank  A. Abolhasani, Y. Cai, E. Ferreira, H. Firouzjahi, 
E. McDonough, G. Moore, O. Ozsoy, Y. Wang and S. Watson for useful discussions. 
This work was supported in
part by a NSERC Discovery grant and by funds from the Canada
Research Chair program (RB).

\end{document}